\documentclass[nofootinbib,a4paper,aps,prd,10pt,superscriptaddress,showkeys,twocolumn]{revtex4}

\usepackage{graphicx}
\usepackage{amsfonts}
\usepackage{amssymb} 
\usepackage{amsmath}
\usepackage{hyperref}
\usepackage{natbib}
\usepackage{float}
\usepackage{dcolumn}

\usepackage{bm}
\usepackage{latexsym,color}

\newcommand{\bea}{\begin{eqnarray}}
\newcommand{\eea}{\end{eqnarray}}
\newcommand{\be}{\begin{equation}}
\newcommand{\ee}{\end{equation}}

\begin{document}

\title{Effects of matter with anisotropic pressure on the Fan-Wang regular black hole shadows}

\author{Yergali~\surname{Kurmanov}}
\email[]{kurmanov.yergali@kaznu.kz}
\affiliation{National Nanotechnology Laboratory of Open Type,  Almaty 050040, Kazakhstan.}
\affiliation{Al-Farabi Kazakh National University, Al-Farabi av. 71,  Almaty 050040, Kazakhstan.}

\author{Orlando~\surname{Luongo}}
\email[]{orlando.luongo@unicam.it}
\affiliation{Al-Farabi Kazakh National University, Al-Farabi av. 71, Almaty 050040, Kazakhstan.}
\affiliation{Universit\`a di Camerino, Divisione di Fisica, Via Madonna delle carceri 9, 62032 Camerino, Italy}
\affiliation{Department of Nanoscale Science and Engineering, University at Albany-SUNY, Albany, New York 12222, USA.}
\affiliation{INAF - Osservatorio Astronomico di Brera, Milano, Italy.}
\affiliation{Istituto Nazionale di Fisica Nucleare (INFN), Sezione di Perugia, Perugia, 06123, Italy.}

\author{Daulet~\surname{Berkimbayev}}
\email[]{daulet9431@mail.ru}
\affiliation{Al-Farabi Kazakh National University, Al-Farabi av. 71,  Almaty 050040, Kazakhstan.}

\author{Kuantay~\surname{Boshkayev}}
\email[]{kuantay@mail.ru}
\affiliation{National Nanotechnology Laboratory of Open Type,  Almaty 050040, Kazakhstan.}
\affiliation{Al-Farabi Kazakh National University, Al-Farabi av. 71, Almaty 050040, Kazakhstan.}
\affiliation{Institute of Nuclear Physics, Ibragimova, 1, Almaty 050032, Kazakhstan.}

\author{Talgar~\surname{Konysbayev}}
\email[] {talgar\_777@mail.ru}
\affiliation{National Nanotechnology Laboratory of Open Type,  Almaty 050040, Kazakhstan.}
\affiliation{Al-Farabi Kazakh National University, Al-Farabi av. 71, Almaty 050040, Kazakhstan.}

\author{Marco~\surname{Muccino}}
\email[]{marco.muccino@unicam.it}
\affiliation{Al-Farabi Kazakh National University, Al-Farabi av. 71, Almaty 050040, Kazakhstan.}
\affiliation{Universit\`a di Camerino, Divisione di Fisica, Via Madonna delle carceri 9, 62032 Camerino, Italy}
\affiliation{Istituto Nazionale di Fisica Nucleare (INFN), Sezione di Perugia, Perugia, 06123, Italy.}
\affiliation{ICRANet, Piazza della Repubblica 10, 65122 Pescara, Italy.}

\author{Guldana~\surname{Rabigulova}}
\email[]{guldanaberikhanovna@gmail.com}
\affiliation{Al-Farabi Kazakh National University, Al-Farabi av. 71,  Almaty 050040, Kazakhstan.}

\begin{abstract}
We here investigate the consequences of an exotic fluid, exhibiting negative radial and tangential pressures, \emph{de facto} violating the Zel'dovich limit, on a regular solution that easily generalizes the Schwarzschild black hole. More precisely, we focus on the regular Fan-Wang spacetime, computing how the black hole shadow images,  surrounded by the quoted fluid, is modified through the presence of \emph{negative} equations of state for the two pressure components. Even though quite different from quintessence, we consider constant radial and tangential equations of state with the aim of emulating, but not reproducing,  dark energy effects. Moreover, we explore the main properties of infalling spherical accretion flows and, accordingly, the influence of the equations of state on the horizons, photosphere, and impact parameter of the Fan-Wang black hole. Afterwards, we examine the luminosities of the shadow and the photon ring in two distinct spherically accretion flows, as well as the observed specific intensity of the shadow itself. Last but not least, we physically interpret the impact of negative pressures on our findings and discuss possible extensions to the isotropic case.
\end{abstract}


\maketitle

\section{Introduction}
\label{sec:Intro}

General relativity (GR) remains the most established theory of gravity, successfully tested by numerous experiments and astronomical observations, like the recent  gravitational wave detection by the LIGO collaboration \cite{2016PhRvL.116f1102A,2016PhRvL.116x1103A,2017PhRvL.119p1101A,2020PhRvD.102d3015A,2021ApJ...915L...5A} and the imaging of black holes (BHs) by the Event Horizon Telescope (EHT) \cite{2019ApJ...875L...1E, 2019ApJ...875L...2E,2019ApJ...875L...3E,2019ApJ...875L...4E,2019ApJ...875L...5E,2019ApJ...875L...6E} that represent a significant breakthrough in the fields of astrophysics and cosmology. In particular, the image of BH accretion presents a dark interior encircled by a luminous ring, referred to as the BH shadow and photon ring, respectively. The photon ring and the shadow are consequences of light bending, or gravitational lensing, caused by BHs as forecasted by GR \cite{1984gere.conf.....W}. Thus, the shadow images of the BHs in M87 and the Milky Way obtained by the EHT provide further compelling evidence for GR. 

However, GR faces several issues,  currently objects of speculation. For example, spacetime singularities cannot account for the quantum nature of BHs \cite{1965PhRvL..14...57P,1970RSPSA.314..529H}, implying the possible existence of regular BHs (RBHs), offering a promising path for the physics beyond GR.
Quite remarkably, the study of RBHs  \cite{Rubio,Bonanno,Borde}, in a manner analogous to singular BHs, encompasses various aspects, including their dynamics \cite{Guo,Yang}, thermodynamic properties \cite{Sharif2020, Wei2018, Huang2023, Yan-Gang,  Lopez}, shadow features \cite{2016PhRvD..93j4004A,2016EPJC...76..630S,2017PhLB..768..373C,2018EPJC...78..399A,2018AnPhy.395..127S,2018GReGr..50..103S,2018CQGra..35w5002A,2018JCAP...12..040H,2019MNRAS.482...52S,2019PhRvD..99d4015H,2019PhRvD.100b4018G,2019PhLB..795....1K,2020EPJC...80..354J,2020PhRvD.101j4001K,2021JCAP...06..037L,2021MNRAS.504.5927A,2021ChPhC..45h5103P,2022ChPhC..46h5106H,2023PhRvD.107f4040X,2024EL....14849001L,2025EPJC...85...46Z,2025JHEAp..4600345Y,2025arXiv250404102L, Zdenek2019,Sau2023,Ling2022, Sushant, Ghosh2020, Ahmed2022, Bambi}, and the characteristics of the accretion disk surrounding RBHs \cite{2025PDU....4801950C,2025NewA..11702354K,kurmanov2024radiative, 2025EPJC...85..514L, 2025IJGMM..2250323A, Zeng, 2024EPJC...84..230B,Guo2023, Narzilloev, Martino2023,Maeda:2021jdc}.

It appears of great importance to explore how a regular solution is embedded with external fields\footnote{For recent developments among novel regular solutions mimicking dark energy at large distances see e.g. \cite{Corona:2024gth,Giambo:2023zmy}.}. Among all, the current cosmological observations show a possible dark energy field, which possesses negative pressure and positive energy density \cite{1998AJ....116.1009R,1999ApJ...517..565P}. Probably the simplest interpretation, beyond the cosmological constant, is the concept of quintessence\footnote{Recent developments have shown that an evolving dark energy fluid cannot be excluded \emph{a priori}, as pointed out by the DESI collaboration, see e.g. \cite{Carloni:2024zpl,Alfano:2024jqn,Luongo:2024fww,Alfano:2024fzv}. }, implying an equation of state $p=\omega\rho$ with $\omega$ the state parameter breaking the Zel'dovich limit, $-1< \omega <-1/3$ \cite{1988ApJ...325L..17P,1998PhRvL..80.1582C}.

Hence, it would be intriguing to explore the effects of quintessence and/or quintessence-like on BH shadows \cite{2024InJPh..98.3019H,2024PhLB..85638963M,2024arXiv241206252G,2025NuPhB101416859M,2025JHEAp..4600350A,2024EPJC...84.1276Y,2022ChPhC..46f5103S}. Particular interest is its study in various geometries and gravitational theories, since additional exotic fluids may be considered in framing the universe dynamics at different stages of its evolution, see e.g. Refs. \cite{Carloni:2025jlk,Dunsby:2023qpb,Dunsby:2016lkw}. 

Nevertheless, quintessence, and/or in general, dark energy fields, require:
\begin{itemize}
    \item[-] the cosmological principle to hold \cite{Li:2011sd}, 
    \item[-] isotropy in the pressures, namely the tangential and radial pressures might be the same \cite{Capozziello:2019cav},
    \item[-] the violation of strong energy condition to drive the universe to accelerate\footnote{For an alternative to the standard $\Lambda$CDM model, see e.g. Refs. \cite{Luongo:2018lgy,Belfiglio:2023rxb}.} \cite{Copeland:2006wr,Luongo:2012dv}.  
\end{itemize}

In this respect, the first static spherically symmetric solution for a BH surrounded by a quintessence-like was proposed by Kiselev \cite{2003CQGra..20.1187K,2003gr.qc.....3031K}, who \emph{did not explicitly} state that the surrounding field is a quintessence one, but something similar to it. Numerous studies followed and considered erroneously that the corresponding field approximates  quintessence. 
However, it was pointed out quite explicitly that \emph{the Kiselev solution is not a solution surrounded by quintessence} \cite{2020CQGra..37d5001V}, as in the original paper \cite{2003CQGra..20.1187K} a terminological error was made, the described effect was incorrectly called ``quintessence'' and, accordingly, Kiselev solution cannot be identified either with an ideal fluid or with quintessence in its classical cosmological understanding as a scalar field with a timelike gradient.

Motivated by the above reasons, in the present study, we examine the properties of the regular and popular solution, dubbed Fan-Wang RBH \cite{Fan:2016hvf}, when enveloped by an anisotropic fluid, with negative equations of state, emulating a quintessence-like configuration, clearly different from quintessence and dark energy\footnote{In our work, we interchange our fluid's name with \emph{quintessence-like} or \emph{exotic fluid}, to indicate a constituent that cannot emulate neither dark energy nor quintessence but whose radial and tangential equations of state for the pressures are constant and negative, at the same time.}. In so doing, we highlight the interplay between gravitational forces and the exotic fluid, breaking the Zel'dovich limit on the equation of state, remarking the main properties near the event horizon. In this study, we explore shadows and rings of Fan-Wang BHs surrounded by spherical accretion, focusing on the influence of the quintessence-like state parameter on the optical appearance of the BHs. We examine the specific intensity of the shadow and the luminosities associated with the quintessence-like Fan-Wang BH shadow and photon ring. Furthermore, we present a comparative analysis of our findings with Schwarzschild, quintessence-like Schwarzschild, and Fan-Wang BHs.

The paper is organized as follows. In Sect. \ref{sec:FW_BH}, we provide a concise overview of Fan-Wang BHs encircled by exotic anisotropic matter, along with an analysis of some of their key properties. In Sect. \ref{sec:FW_Light}, we focus on photon trajectories in the spacetime of Fan-Wang BHs influenced by exotic anisotropic matter, and examine how the model parameters affect their behavior. Sect. \ref{sec:Shadow} is devoted to the investigation of shadow depictions of quintessence-like Fan-Wang BHs under static and spherically symmetric accretion flow conditions.  Finally, in Sect. \ref{Conclusion}, we present the conclusions and perspectives of our work.

\section{Fan-Wang metric surrounded by exotic anisotropic matter} \label{sec:FW_BH}

The Fan-Wang RBH is expressed as \cite{FanWang,2022JHEP...11..108M,2024PDU....4601566K,2025arXiv250318380S}
\begin{equation}\label{metr_generic}
    ds^2=-f(r)dt^2+\frac{dr^2}{f(r)}+r^2(d\theta^2+\sin^2{\theta}d\phi^2),
\end{equation}
with the function
\begin{equation}
    f(r)=1-\frac{2Mr^2}{(r+l)^3}.
\end{equation}
where $l$ is the magnetic charge parameter for the Fan-Wang BH and $M$ represents the mass of BH. The event horizons exist only when $l\leq8/27$ \cite{2024PDU....4601566K}, whereas for $l=0$ the Schwarzschild solution is recovered. 

In Ref.~\cite{2003CQGra..20.1187K},  a novel metric solution for a stationary, spherically symmetric spacetime, taking into account that the energy-momentum tensor of quintessence-like has been proposed \cite{2016PhRvD..94j6005C}
\begin{equation}
   T_{\phi}^{\phi}=T_{\theta}^{\theta}=-\frac{1}{2}\left(3\omega +1\right)T_{r}^{r}=\frac{1}{2}\left(3\omega +1\right)T_{t}^{t}
\end{equation}
where $\omega$ corresponds to the equation of state of the exotic fluid     \emph{that is not quintessence as it is anisotropic in pressures}. 

The energy density, $\rho= T_{tt}\geq0$, with the strong energy condition,  $\left|3\omega + 1\right|\leq 2$. Using these principles, the metric function for a BH surrounded by the exotic fluid is derived by incorporating the term $-a/r^{3w+1}$ into the metric of the BH \cite{2012GReGr..44.1857F,2017Ap&SS.362..218G}, yielding 
\begin{equation}\label{FWquin}
     f(r)=1-\frac{2Mr^2}{(r+l)^3}-\frac{a}{r^{3\omega+1}}.
\end{equation}
where  $a$ acts as a normalization factor, which is a positive constant utilized in the metric equation, and the range of our exotic fluid\footnote{In the situation corresponding to $a = 0$, the aforementioned metric reduces to the Fan-Wang BH. In the context of $l=0$, it delineates a Schwarzschild BH surrounded by exotic anisotropic matter, as initially derived by Kiselev in Ref.~\cite{2003CQGra..20.1187K}. Additionally, for $a=l=0$, the result is a Schwarzschild BH, which is characterized by a single event horizon.},
\begin{equation}
\omega\in(-1;-1/3).  
\end{equation}

In this work, we examine $\omega=-0.5$ and $\omega=-0.7$, that produce sufficiently noticeable differences in the behavior of the horizons, photon sphere, shadow radius, and radiation intensity. This enables a clear illustration of how the parameter $\omega$ affects the geometry and optical properties of BHs. Values such as $\omega=-0.4$ are too close to the boundary $\omega=-1/3$, where the effect of exotic anisotropic matter becomes weak. Further, values such as $\omega=-0.8$ appear too close to the cosmological constant case, i.e.,  $\omega=-1$. Subsequently, we derive the necessary conditions under which the Fan-Wang spacetime, when enveloped by exotic matter, may possess three horizons. By setting the equation $f(r)=0$, one can elucidate the relationship between the BH mass and the horizon radius as follows
\begin{equation}
    M(r)=\frac{(r+l)^3}{2r^2} \left(1- \frac{a}{r^{3\omega+1}}\right).
\end{equation}

\begin{figure}
\includegraphics[width=1\linewidth]{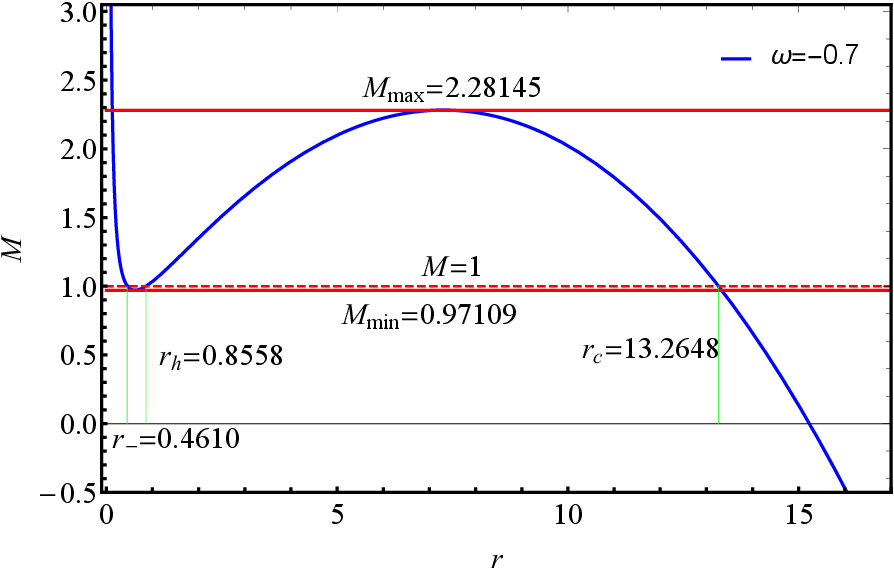}
\caption{The mass of the BH as a function of the radial coordinate $r$ for parameters $l=8/27$, $a=0.05$, and $\omega=-0.7$.}
\label{fig:BHMass}
\end{figure}

The function $M$ is depicted in Fig. \ref{fig:BHMass} for selected and appropriate values of the constants $l$, $a$, and $\omega$. 
It is evident that this function exhibits both a local maximum and a minimum.  When the particular values of $M$ lie within the interval defined by $M_{min}<M<M_{max}$, the BH manifests three distinct horizons. The inner horizon is designated as $r_{-}$, the event horizon is designated as $r_{h}$, and the cosmological horizon is designated as $r_{c}$. Moreover, there exists a pivotal value for the normalization factor $a$, at which $r_{-}$,  $r_{h}$, and $r_{c}$ coincide ($r_{-}=  r_{h}=r_{c}$). This particular critical value, denoted as $a_{crit}$, is ascertained as follows:
\begin{eqnarray}\label{acrit}
   a_{crit}&=&\frac{1}{3\omega\left(3\omega-\sqrt{4+4\omega+9\omega^{2}}\right)}\nonumber\\
   &\times& 
   \Bigg[2^{-1-3\omega}\bigg(\frac{l\left(-2+\omega-\sqrt{4+4\omega+9\omega^{2}}\right)}{\omega}\bigg)^{1+3\omega}\nonumber\\
   &\times& \left(2+3\omega+\sqrt{4+4\omega+9\omega^{2}}\right)\Bigg].
\end{eqnarray}

For $a>a_{crit}$, no Fan-Wang solution is available for any values of $M$, regardless of the specified values of $M$. Conversely, for $a<a_{crit}$, Fan-Wang solutions are attainable within the range of $M_{min}<M<M_{max}$.

The values of $M_{min}$ and $M_{max}$ for $a = 0.05$, $\omega=-0.7$, and various values of $l$ are displayed in Tab. \ref{tab:Mminmax}.  Additionally, from Eq. (\ref{acrit}), the critical value of $a$ is determined for $\omega=-0.7$ across different values of $l$. In this work, we set $a=0.05$ and $M=1$, which falls within the range of $M_{min}$ and $M_{max}$. Consequently, the value of $a$ remains less than the critical value for all considered values of $l$, thereby resulting in BH solutions characterized by three distinct horizons.

\begin{table}[ht]
\begin{center}
\caption{The values of $a_{crit}$, $M_{min}$, $M_{max}$ for $a=0.05$, $\omega=-0.7$ and various values of $l$}
\vspace{3 mm}
\label{tab:Mminmax}
\begin{tabular}{ccccc}
\hline
\hline
$ \omega $  &  $l$   & $a_{crit}$        & $M_{min}$        & $M_{max}$  \\
       
\hline
-0.7 &     2/27      &  1.02825    & 0.24821  & 2.09133 \\   
-0.7 &    4/27        &  0.47969   & 0.49287  & 2.15245 \\
-0.7 &    6/27       &   0.30709   & 0.73361  & 2.21569 \\
-0.7&     8/27        &  0.22379   & 0.97016  & 2.28118 \\
\hline
\hline
  \end{tabular}
  \end{center}
\end{table}

\begin{figure}
\includegraphics[width=1\linewidth]{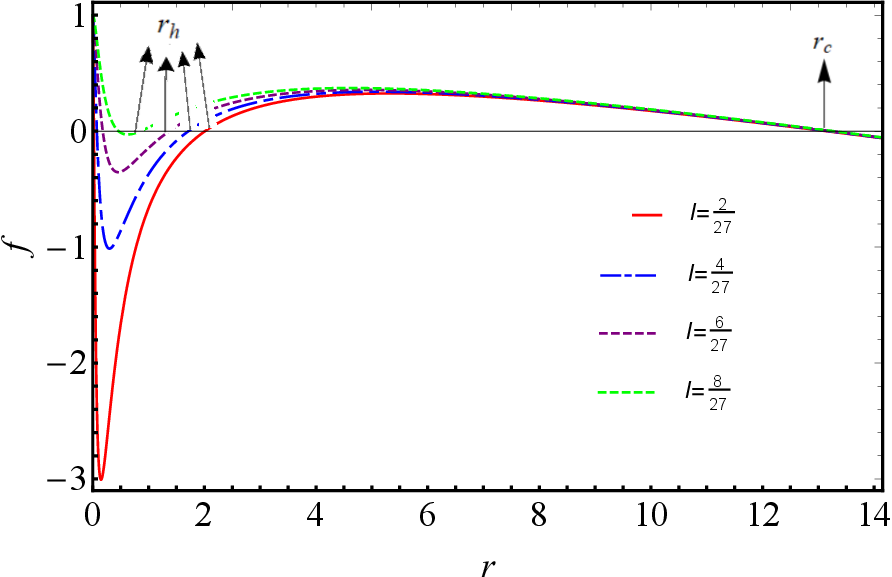}
\caption{The metric coefficient $f(r)$ as a radial coordinate $r$ at $M$ = 1, $a$ = 0.05 and $\omega$ =-0.7 with varying values of $l$.}
\label{fig:fomega}
\end{figure}

Fig. \ref{fig:fomega} depicts the behavior of the function $f(r)$ with $M=1$, $a = 0.05$, and $\omega=-0.7$. The $f(r)$ function has three real roots, which correspond to $r_{-}$, $r_{h}$, and $r_{c}$. Tab. \ref{tab:Horizon} shows that increasing the absolute value of the $\omega$ parameter reduces the difference between the event and cosmological horizons.

\section{Light Deflection} \label{sec:FW_Light}

To comprehend how light behaves in proximity of our BH configuration, we work out trajectories of light rays. The Euler-Lagrange equation yields 
\begin{equation}
\frac{d}{ds}\left(\frac{\partial\mathcal{L}}{\partial\dot{x}^{i}}\right) = \frac{\partial\mathcal{L}}{\partial x^{i}},
\label{eq:euler-lagrange}
\end{equation}
where the sign “$\dot{\,\,}$” denotes the derivative with respect to $s$, and the variable  $s$  represents the affine parameter governing the light path.  
Moreover, $x^i$ is the light ray's four-velocity in this case, and $\mathcal{L}$  is the Lagrangian density, 
\begin{eqnarray}
\mathcal{L} &=& \frac{1}{2} g_{\mu\nu} \dot{x}^{\mu} \dot{x}^{\nu}\nonumber\\ &=& \frac{1}{2} \left( - f(r) \dot{t}^2 + \frac{\dot{r}^2}{f(r)}  + r^2 \dot{\theta}^2 + r^2 \sin^2 \theta \dot{\varphi}^2 \right).
\label{eq:Lagrangian}
\end{eqnarray}
The light rays will always move within the equatorial plane if we specify certain initial conditions, specifically $\theta=\frac{\pi}{2}$ and $\dot{\theta}=0$. 

The photon energy $E$ and its angular momentum $L$, in the time and azimuthal directions, are both conserved quantities. Combining Eqs.~(\ref{FWquin}), (\ref{eq:euler-lagrange}), and (\ref{eq:Lagrangian}), one finds
\begin{subequations}
\begin{align}
&\dot{t} = - \frac{E}{1 - \frac{2Mr^2}{(r + l)^3} - \frac{a}{r^{3\omega+1}}},\label{eq:time_derivative}\\
&\dot{\varphi} = \frac{L}{r^2},
\label{eq:phidot}\\
&\dot{r}^2 + \left( 1 - \frac{2Mr^2}{(r + l)^3} - \frac{a}{r^{3\omega+1}} \right) \left( \frac{L^2}{r^2} + h \right) = E^2.
\label{eq:energy_relation}
\end{align}
\end{subequations}

For null geodesics, $h=0$, the kinematics and effective potential, respectively, read
\begin{subequations}
\begin{align}
&\dot{r}^2 +\frac{L^2}{r^2} \left( 1 - \frac{2Mr^2}{(r + l)^3} - \frac{a}{r^{3\omega+1}} \right) = E^2,
\label{eq:energy_relation2}\\
&V_{\mathrm{eff}}(r) = \frac{L^{2}}{r^{2}} f(r) = \frac{L^{2}}{r^{2}} \left(1 - \frac{2Mr^{2}}{(r + l)^{3}} - \frac{a}{r^{3\omega + 1}}\right).
\label{eq:effective_potential}
\end{align}
\end{subequations}

\begin{figure*}[ht]
\begin{minipage}{0.49\linewidth}
\center{\includegraphics[width=0.99\linewidth]{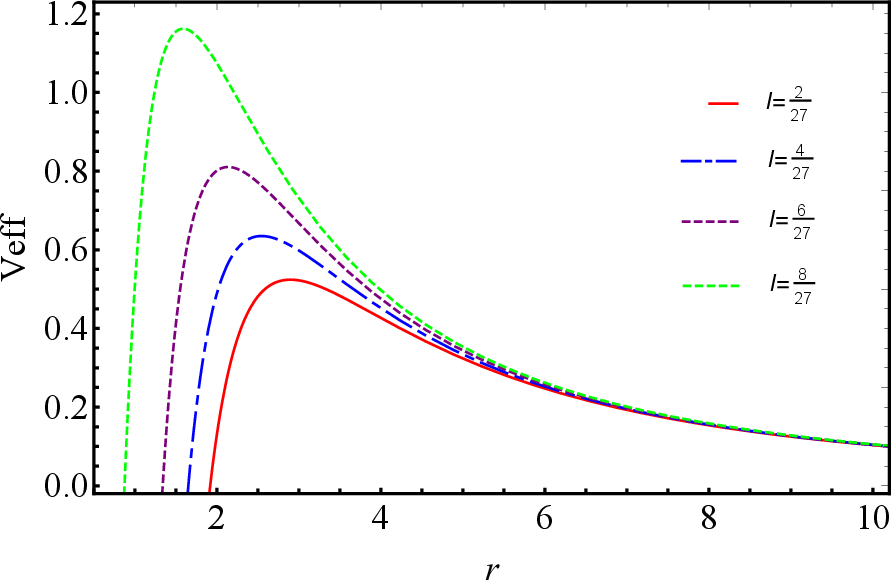}\\ }
\end{minipage}
\hfill 
\begin{minipage}{0.50\linewidth}
\center{\includegraphics[width=0.98\linewidth]{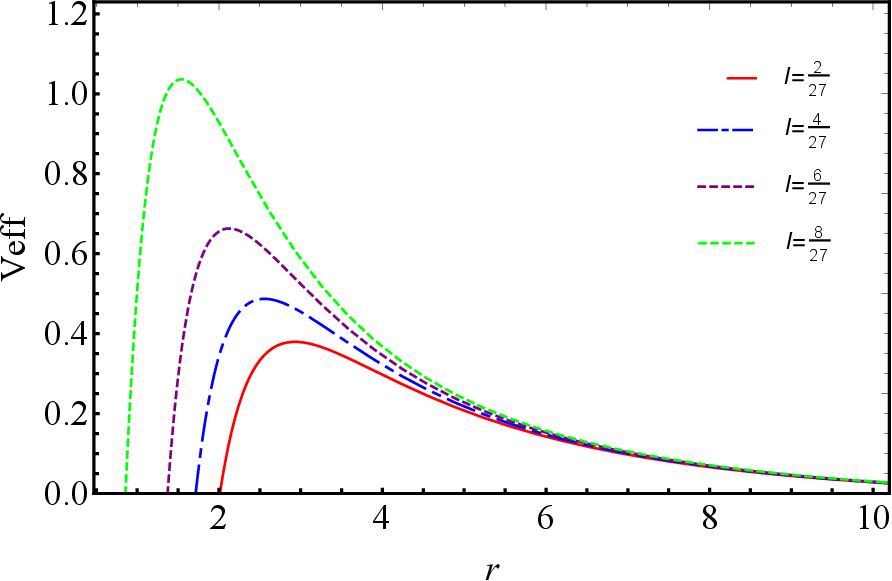}\\ }
\end{minipage}
\caption{The variation of the effective potential with respect to the radial coordinate $r$  for various values of $l$ at $a = 0.05$, $\omega=-0.5$ (left panel) and $\omega=-0.7$ (right panel).}
\label{fig:potentialFW}
\end{figure*}

\begin{figure}
\includegraphics[width=1\linewidth]{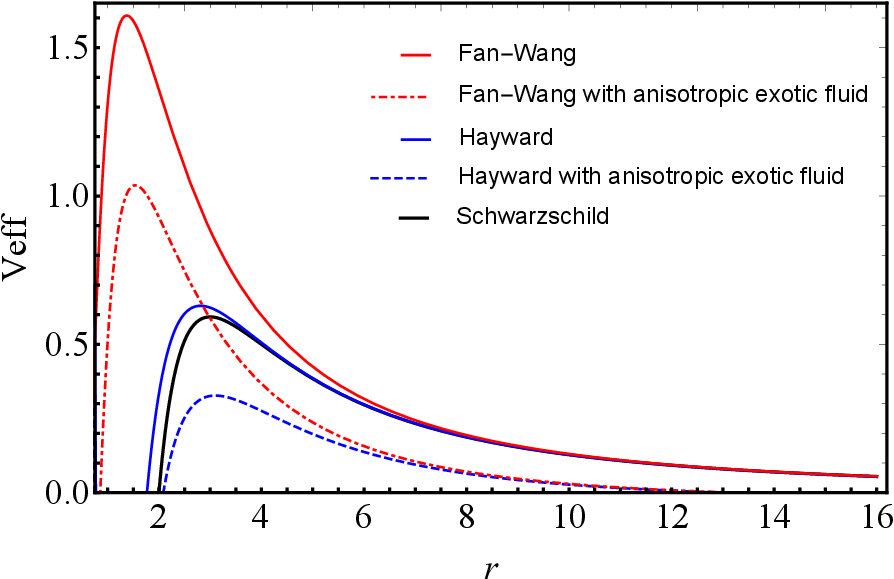}
\caption{The effective potential $V_{eff}$ as a function of the radial coordinate $r$ for Schwarzschild space-time, Hayward BH with $g$=0.9, $a$=0, Hayward BH surrounded by anisotropic exotic fluid with $g$ = 0.9, $a$ = 0.05, $\omega$ =-0.7, $M$ = 1 (Reproduced from \cite{2024InJPh..98.3019H}), Fan-Wang BH at $l$ = 8/27 and Fan-Wang BH surrounded by anisotropic exotic fluid with $a$=0.05, $\omega$ =-0.7 and $M$ =1 at $l$ = 8/27.}
\label{fig:potential}
\end{figure}

Fig. \ref{fig:potentialFW} displays the effective potential, modified by the presence of our exotic fluid, for a range of $l$ parameters. In the left and right panels, the state parameter is configured to $\omega=-0.5$ and $\omega=-0.7$, respectively. It is evident that an enhancement in the parameter $l$ results in an elevated peak of the potential, whereas an augmentation in the absolute value of $\omega$ causes a reduction in the maximum value of the potential.

The influence of the exotic fluid  on the effective potential is illustrated in Fig. \ref{fig:potential}. For comparative purposes, the effective potential for Hayward, both with and without the influence of the source fluid was likewise reproduced from Ref. \cite{2024InJPh..98.3019H}. The results demonstrate that the presence of the external fluid  results into a decreased maximum of the effective potential. 
Additionally, for all values of $l$, the peak of the effective potential, in the presence of external fluid,  associated with Hayward BHs, is lower than that of Fan-Wang, Hayward, and Schwarzschild BHs.

Utilizing a ray-tracing methodology, we ascertain the characteristics of the photosphere, notably its location at $r=r_{ph}$, and delineate the associated impact parameter. Eq. (\ref{eq:energy_relation}) serves as the criterion for the identification of orbits associated with the photosphere
\begin{equation}
V_{\mathrm{eff}}(r_{ph}) = E_{ph}^{2}, \quad V_{\mathrm{eff}}'(r_{ph}) = 0,
\label{eq:photon_sphere_conditions}
\end{equation}
here the prime denotes differentiation in relation to the coordinate $r$ on the radial. Utilizing Eq. (\ref{eq:effective_potential}) results in the subsequent relationship
\begin{equation}
rf'(r) - 2f(r) = 0.
\label{eq:critical_condition}
\end{equation}

The equation for the photosphere that results from substituting $f(r)$ from Eq. (\ref{FWquin}) is
\begin{equation}
6 M r^3 - 2 \left( l + r \right)^4 + 3a r^{-1-3\omega} \left( 1 + \omega\right) \left( l+ r \right)^4 = 0,
\label{eq:system_equation}
\end{equation}

The roots of Eq.~(\ref{eq:system_equation}) are obtained numerically. In the Schwarzschild metric, the photon radius is determined to be $r_{ph} =3M$ for $a=l=0$.  Furthermore, when considering the case where the parameters $a=0$, $l\neq0$, corresponding to Fan-Wang spacetime, the aforementioned equation lacks an analytical solution. Conversely, for Schwarzschild BHs with quintessence-like matter at $l=0$, $a\neq0$, an exact solution is obtainable solely for $\omega= -2/3$, as demonstrated below
\begin{equation}
r_{ph} = \frac{1 - \sqrt{1 - 6Ma}}{a},
\label{eq:photon_sphere_radius}
\end{equation}

Furthermore, the impact parameter of the photosphere $b_{ph}=L_{ph}/E_{ph}$ is articulated as follows
\begin{equation}
b_{ph} = \frac{r_{ph}}{\sqrt{f(r_{ph})}}.
\label{eq:critical_impact_param}
\end{equation}

Tab. \ref{tab:Horizon} presents the numerical outcomes of the inner horizon radius $r_{-}$, the event horizon radius $r_{h}$, the cosmological horizon $r_{c}$, the radius of the photosphere $r_{ph}$, and the impact parameter of the photosphere $b_{ph}$ for varying parameters of $l$ and  $\omega$ with $M=1$ and $a=0.05$.  In the analysis of the shadow Fan-Wang BH within a quintessence-like framework, the cosmological horizon is of significant importance. This is due to the consideration of a distant observer situated proximate to the cosmological horizon and accretion flux in proximity to the event horizon. The disparity between the inner horizon radius $r_{-}$ and the event horizon radius $r_{ph}$ is contingent upon the values of the state parameters $\omega$ and $l$. As can be seen from Tab. \ref{tab:Horizon}  that increasing the absolute value of $\omega$ leads to a decrease in this difference. In addition, when $\omega$ is fixed and the parameter $l$ increases, the parameters $r_{h}$, $r_{ph}$ and $b_{ph}$ decrease. It is also noted that when $l$ is fixed, with an increase in the absolute value of $\omega$ results in an increase in the values of $r_{h}$ and $b_{ph}$ are increase, while the cosmological horizon $r_{c}$ exhibits a reduction. It is worth noting that the data presented in the first row of each entry in $\omega$ from Tab. \ref{tab:Horizon} are related to the Schwarzschild quintessence-like BH.

\begin{table}[ht]
\begin{center}
\caption{The numerical values corresponding to the inner horizon $r_{-}$, event horizon $r_{h}$, cosmological horizon $r_{c}$, photon radius $r_{ph}$, and impact parameter $b_{ph}$ are presented for varying values $l$ and $\omega$, with conditions $M=1$ and $a=0.05$.}.
\vspace{3 mm}
\label{tab:Horizon}
\begin{tabular}{ccccccc}
\hline
\hline
$ \omega $  &  $l$   & $r_{-}/M$        & $r_{h}/M$        & $r_{c}/M$  & $r_{ph}/M$        & $b_{ph}/M$ \\
       
\hline
$-0.4$ &        $0$     &      -   & $2.1234$ & $319999$ & $3.1804 $ & $5.7295$\\   
       &     $2/27$      &  $0.0202 $   & $1.8894 $ & $3.2\times 10^{6}$ & $2.8685 $ & $5.3049$\\   
       &   $ 4/27$        &  $0.0719 $  & $1.6307 $ & $3.2\times 10^{6}$ & $2.5250 $ & $4.8390$\\
       &    $6/27 $      &   $0.1718  $ & $1.3272 $ & $3.2\times 10^{6}$ & $2.1276 $& 4.$3064$\\
       &    $ 8/27$        & $0.4166 $   & $0.8812$  &$ 3.2\times 10^{6} $& $1.5987$  & $3.6305$\\
\hline
$-0.5$ &     $0$     &      -   & $2.1586$ & $395.969$ & $3.2163 $ & $5.9881$\\   
       &   $2/27$      &  $0.0205$   & $1.9179$  & $395.972$ & $2.8967$  & $5.5255$\\   
       &   $4/27$        &  $0.0731$   & $1.6518$  & $395.974$ &$ 2.5445$ & $5.0196$\\
       &   $6/27$       &   $0.1755$   & $1.3395 $ & $395.976$ &$ 2.1365$ & $4.4429$\\
       &   $8/27$        &  $0.4337$  & $0.8736 $ & $395.979$ & $1.5890$  & $3.7107$\\
\hline
$-0.6$ &     $0$     &      -   & $2.2080$ & $39.6448$ & $3.2504 $ & $6.4204$\\   
    &     $2/27$      &  $0.0206$   & $1.9563$  & $39.6605$ & $2.9205$  & $5.8775$\\   
       &    $4/27$        &  $0.0736$   & $1.6789$  & $39.6761$ &$ 2.5576$ & $5.2920$\\
       &   $6/27$       &   $0.1777$   & $1.3540 $ & $39.6916$ &$ 2.1375$ & $4.6342$\\
      &   $8/27$        &  $0.4484$  & $0.8651 $ & $39.7069$ & $1.5700$  & $3.8090$\\

\hline
$-0.7$ &     $0$     &      -   & $2.2830$ & $13.1027$ & $3.2710 $ & $7.2343$\\   
       &     $2/27$      &  $0.0206$   & $2.0112$  & $13.1454$ & $2.9299$  & $6.4945$\\   
       &    $4/27$        &  $0.0739$   & $1.7151$  & $13.1866$ &$ 2.5562$ & $5.7320$\\
       &   $6/27$       &   $0.1791$   & $1.3719 $ & $13.2264$ &$ 2.1247$ & $4.9132$\\
      &   $8/27$        &  $0.4610$  & $0.8558$ & $13.2648$ & $1.5390$  & $3.9297$\\

\hline
$-0.8$ &     $0$     &      -   & $2.4149$ & $6.5509$ & $3.2548 $ & $9.2206$\\  
       &     $2/27$      &  $0.0206$   & $2.0985$  & $6.6455$ & $2.9070$  & $7.7890$\\   
       &    $4/27$        &  $0.0740$   & $1.7668$  & $6.7317$ &$ 2.5276$ & $6.5286$\\
       &   $6/27$       &   $0.1799$   & $1.3945 $ & $6.8085$ &$ 2.0911$ & $5.3438$\\
      &   $8/27$        &  $0.4721$  & $0.8456$ & $6.8784$ & $1.4946$  & $4.0763$\\

\hline
\hline
  \end{tabular}
  \end{center}
\end{table}

Comparison of the crucial parameters $r_{h}$, $r_{c}$, $r_{ph}$, and $b_{ph}$ is provided in Tab. \ref{tab:HorizonII}. The first row represents the outcomes for the Schwarzschild BH. The numerical values corresponding to the Fan-Wang BH are presented in the second row. The subsequent rows, specifically the third and fourth, delineate the parameters for the Schwarzschild space-time and Fan-Wang space-time in the presence of anisotropic exotic fluid, respectively. The key finding from Tab. \ref{tab:HorizonII} is that the differences in the sizes of horizons, photon spheres, and impact parameters between the Fan-Wang and Fan-Wang with anisotropic exotic fluid are comparatively insignificant.

The differences in the photon radius and impact parameter between the Fan-Wang BH and quintessence-like Fan-Wang BH can be quantitatively estimated. In particular, the event horizon radius increases by approximately $13.3\%$, the photon sphere radius by $9.4\%$, and the impact parameter bph by about $31.7\%$. Despite the influence of anisotropic matter on the photon radius and impact parameter, the resulting changes are within a moderate range and do not lead to substantial alterations in the BH’s shadow.

\begin{table}[ht]
\begin{center}
\caption{The values of the event horizon $r_{h}$, cosmological horizon $r_{c}$, photon radius $r_{ph}$, and impact parameter $b_{ph}$ for Schwarzschild, Fan-Wang, both Schwarzschild and Fan-Wang BHs when surrounded by exotic anisotropic matter.}
\vspace{3 mm}
\label{tab:HorizonII}
\begin{tabular}{cccccccc}
\hline
\hline
BH type & $\omega $  &  $a$   & $l$        & $r_{h}/M$        & $r_{c}/M$  & $r_{ph}/M$        & $b_{ph}/M$ \\
       
\hline
Schwarzschild & - &     $0$     &     $0$   & $2$ & - & $3$ & $5.20$\\  
\hline
Fan-Wang &   -    &  $0$   & $4/27$  & $1.51$ & - &$ 2.34$ & $ 4.35$\\ 
\hline
quintessence-like\\
Schwarzschild   &   $-0.7$ &  $0.05$   & $0$  & $2.28$ & $13.10$ & $3.27$ & $7.23$\\

\hline
 quintessence-like\\
 Fan-Wang   & $-0.7$  &   $0.05$       &   $4/27$   & $1.71 $ & $13.19$ &$2.56$ & $5.73$\\
    
\hline
\hline
  \end{tabular}
  \end{center}
\end{table}

By employing Eqs. (\ref{eq:phidot}) and (\ref{eq:energy_relation}) along with the substitution $u=1/r$, we obtain
\begin{equation}
\frac{du}{d\varphi} = \sqrt{\frac{1}{b^2} - u^2 \left(1 - \frac{2M}{u^{2}\left(\frac{1}{u} + l\right)^3} - a u^{3\omega + 1}\right)}.
\label{eq:photon_trajectory}
\end{equation}

\begin{figure}
\includegraphics[width=1\linewidth]{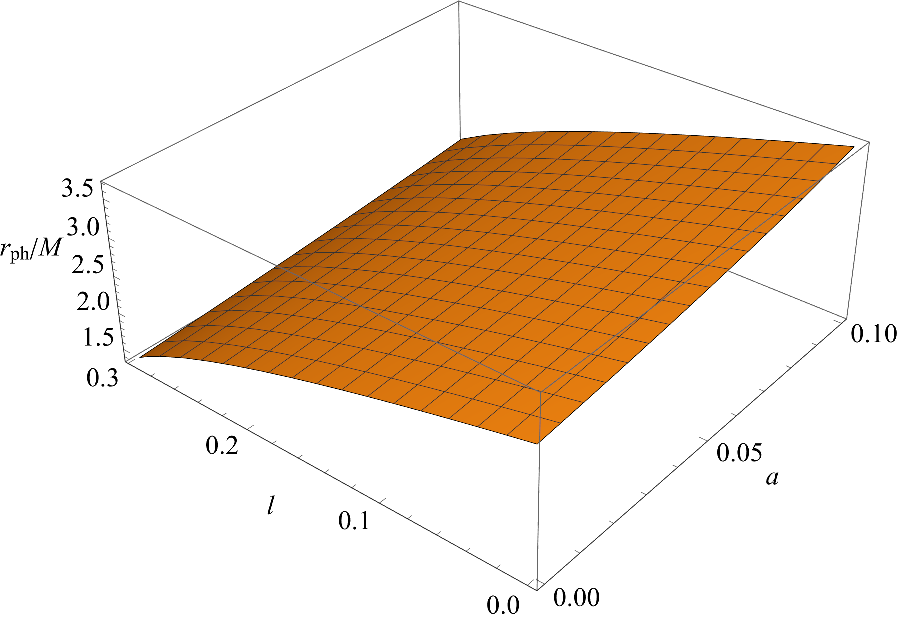}
\caption{The dependence of the photon radius on the parameters $l$ and $a$ for $\omega=-\frac{2}{3}$.}
\label{fig:PhRadius}
\end{figure}

Fig.~\ref{fig:PhRadius} shows the dependency of the photon radius on the parameters $a$ and $l$ at $\omega=-2/3$ for the Fan-Wang BH with anisotropic exotic fluid. The figure illustrates that the photon radius diminishes for increasing values of $l$, given a fixed $a$. Conversely, for a fixed value of $l$, the photon radius expands with the enhancement of $a$.

\begin{figure*}[ht]
\begin{minipage}{0.49\linewidth}
\center{\includegraphics[width=0.99\linewidth]{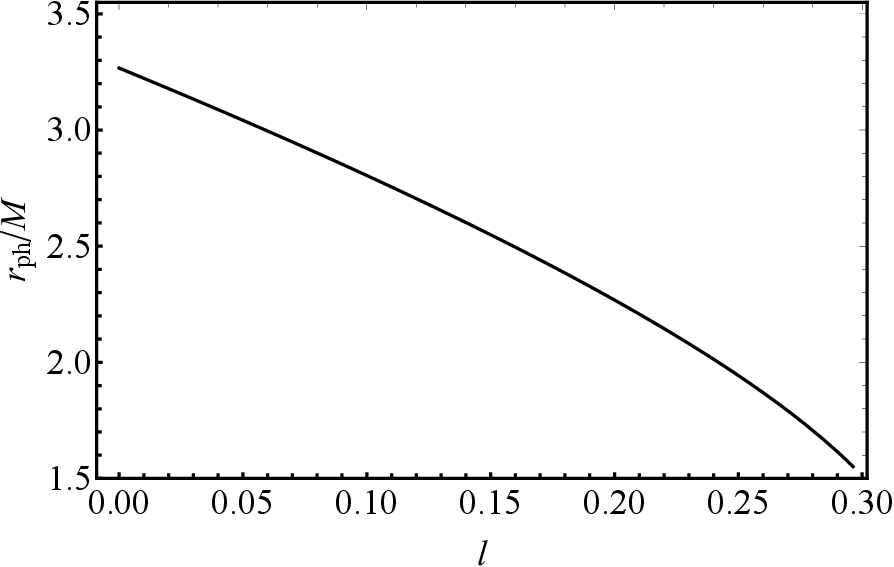}\\ }
\end{minipage}
\hfill 
\begin{minipage}{0.50\linewidth}
\center{\includegraphics[width=0.98\linewidth]{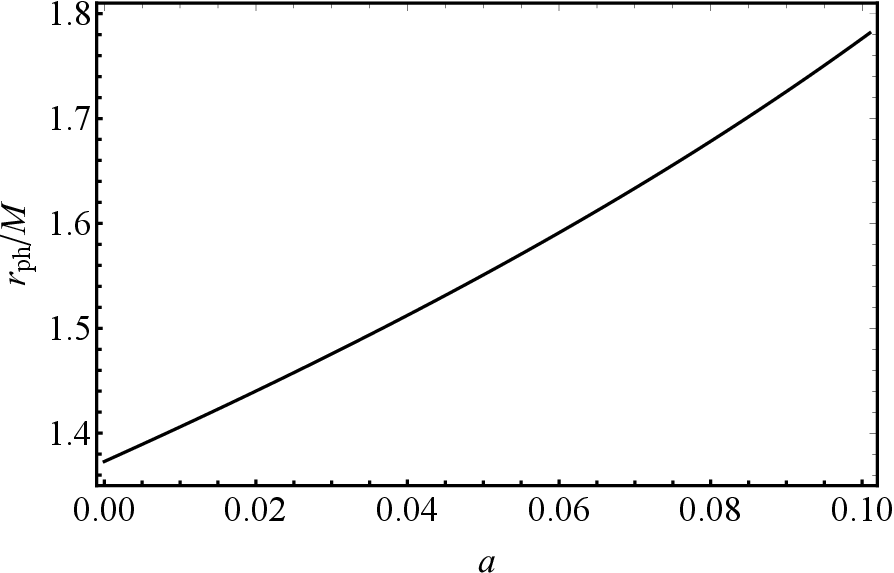}\\ }
\end{minipage}
\caption{Photon sphere radius as a function of parameter $l$ for $a=0.05$ (left panel) and parameter a for $l=8/27$ (right panel) at fixed $\omega = -\frac{2}{3}$.}
\label{fig:PhRadiusLR}
\end{figure*}

Fig.~\ref{fig:PhRadiusLR} shows the photon sphere radius as a function of the parameters $l$ and $a$, with a fixed value of $\omega = -\frac{2}{3}$. The photon sphere radius is plotted as a function of parameter $l$ at fixed $a = 0.05$ (left panel), and as a function of parameter $a$ at fixed $l = \frac{8}{27}$ (right panel). This two-panel format allows for a clear analysis of the influence of each parameter.

\begin{figure*}[ht]
\begin{minipage}{0.49\linewidth}
\center{\includegraphics[width=0.99\linewidth]{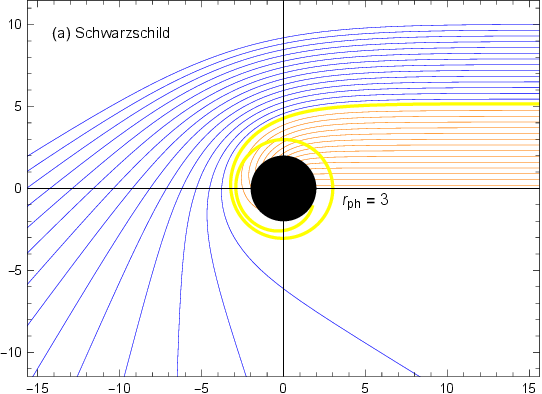}\\ }
\end{minipage}
\hfill 
\begin{minipage}{0.50\linewidth}
\center{\includegraphics[width=0.98\linewidth]{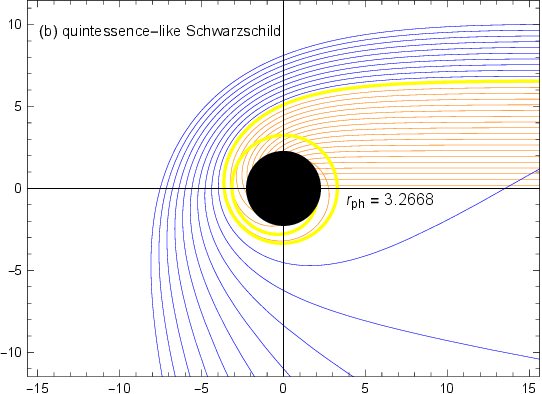}\\ }
\end{minipage}
\begin{minipage}{0.49\linewidth}
\center{\includegraphics[width=0.99\linewidth]{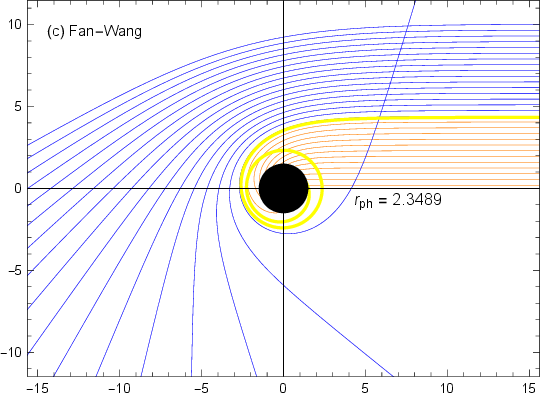}\\ }
\end{minipage}
\begin{minipage}{0.50\linewidth}
\center{\includegraphics[width=0.98\linewidth]{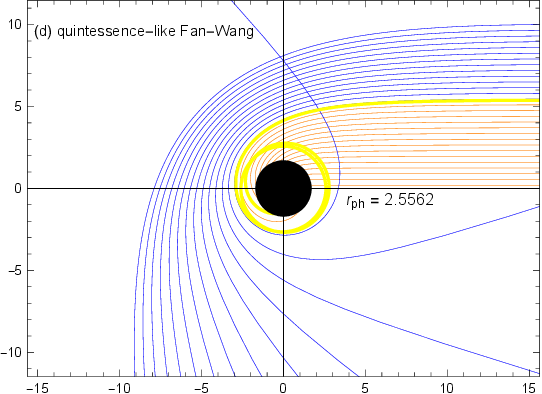}\\ }
\end{minipage}
\caption{The polar plots depict the paths of light rays surrounding a Schwarzschild BH with and without the presence of quintessence-like matter on the top row, and a Fan-Wang BH with and without quintessence-like matter on the bottom row, given $l=4/27$ and $a = 0.05$, $\omega = -0.7$ at $M=1$. The yellow, blue, and orange lines illustrate the trajectories of light rays where $E = E_{ph}$, $E < E_{ph}$, and $E > E_{ph}$, respectively.}
\label{fig:Lightrays}
\end{figure*}

Fig.~\ref{fig:Lightrays} shows that the shadow and photosphere are created by light deflection, assuming that light rays approach the BH from the right side. Panels (a),(b),(c), and (d) illustrate the light trajectories within the metrics of Schwarzschild, quintessence-like Schwarzschild, Fan-Wang, and quintessence-like Fan-Wang spacetimes. The results for Schwarzschild and quintessence-like Schwarzschild spacetimes are displayed in the top row, whereas the bottom row illustrates the findings for Fan-Wang BHs both without and with the presence of quintessence-like matter. By examining the outcomes of figures (a) with (b), alongside figures (c) with (d), it becomes evident that quintessence-like matter leads to an increased deflection of light rays \cite{2020AnPhy.41868183J}. 

The apparent size of the BH shadow as observed from a distance is measured by the angular diameter \cite{2022PhR...947....1P}, one particular measure associated with the shadow.  The angular diameter is expressed as follows
\begin{equation}
\Omega = \frac{2 b_{ph}}{D},
\label{eq:omega_definition}
\end{equation} 
where $D$ is the distance of the BH to the observer. The following is another way to express Eq.~\eqref{eq:omega_definition}
\begin{equation}
\left( \frac{\Omega}{\mu\text{as}} \right) = \left( \frac{6.191165 \times 10^{-8}}{\pi} \frac{M/M_\odot}{D/\text{Mpc}} \right) \left( \frac{b_{ph}}{M} \right),
\label{eq:omega_relation}
\end{equation}
where $b_{ph}$ is defined by Eq. (\ref{eq:critical_impact_param}).

\begin{figure*}
\begin{minipage}{0.49\linewidth}
\centering
\includegraphics[width=\linewidth]{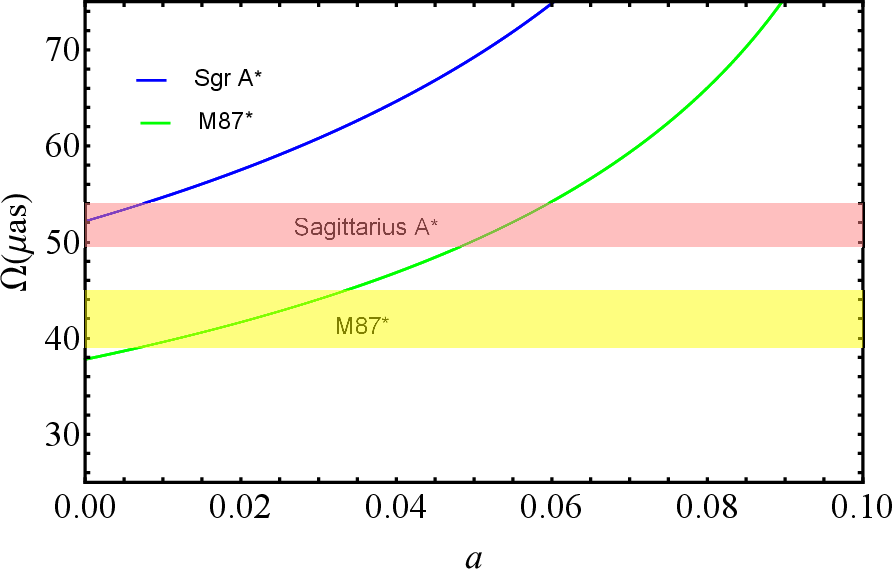}
\end{minipage}
\hfill 
\begin{minipage}{0.49\linewidth}
\centering
\includegraphics[width=\linewidth]{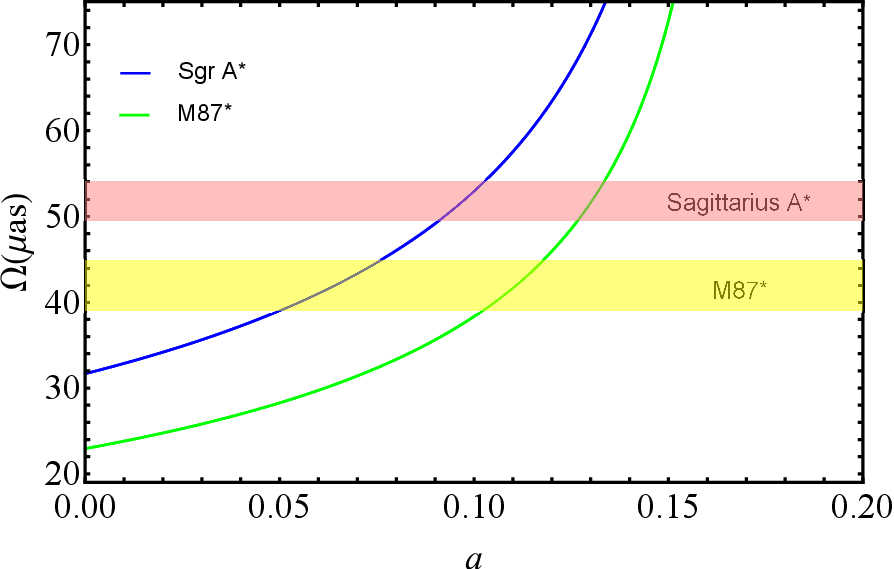}
\end{minipage}
\caption{The relationship between the parameter $a$  and the observed shadow's angular diameter for Schwarzschild (left panel, Reproduced from Ref. \cite{2024InJPh..98.3019H}) and Fan-Wang (right panel) BHs with quintessence-like matter at $8/27$ for $\omega=-2/3$. }
\label{fig:Angulardiameter}
\end{figure*}

By employing the shadow diameter determined through EHT observations, it becomes feasible to ascertain constraints on the free parameters associated with quintessence-like Schwarzschild and Fan-Wang with $l=8/27$. Fig. \ref{fig:Angulardiameter} presents an examination of the quintessence-like Schwarzschild and Fan-Wang BHs, with $\omega=-\frac{2}{3}$. The pink and yellow regions show Sagittarius A* ($51.8\pm 2.3  \mu as$) and M87* ($42 \pm 3 \mu as$) experimental data. According to current observations Refs. \cite{EventHorizonTelescope:2019dse,EventHorizonTelescope:2022wkp}, Sagittarius A* corresponds to $y = 4.14 \times10^{6}$ and distance $D = 8.127$ kpc, while $M87^{*}$ corresponds to $y=6.2 \times 10^{9}$ and distance $D = 16.8$ Mpc. As a result, we restrict the parameter $a$ of the quintessence-like Fan-Wang BH to the range $0.102<a<0.118$ for $M87^{*}$, and $0.090<a<0.103$ for Sagittarius A*. For the Schwarzschild with quintessence-like matter, the value of $a$ is provided in Ref. \cite{2024InJPh..98.3019H}.

\section{Observing BH Shadows with Spherical  Accretions} \label{sec:Shadow}

This section is dedicated to exploring the visual properties of Fan-Wang BHs enveloped by quintessence-like matter. Our primary aim is to determine the extent to which these properties are affected by two significant parameters $l$ and the quintessence-like state parameter $\omega$. To achieve this objective, we start examining a static model of spherical accretion, characterized by accretion flows that are both optically and geometrically thin. Subsequently, our attention will transition to dynamic scenarios that entail infalling spherical accretion flows.

\subsection{The static model of spherical accretion}

The objective of our study is to examine the shadow and photosphere of a Fan-Wang enveloped by quintessence-like matter, within the framework of static spherical accretion. Our principal focus is on the analysis of the specific intensity as observed by an observer, which is customarily quantified in units of ($\text{erg}\, \text{s}^{-1}\, \text{cm}^{-2}\, \text{str}^{-1}\, \text{Hz}^{-1}$)  \cite{1997A&A...326..419J, 2013PhRvD..87j7501B}  and it can be stated in this way
\begin{equation}
I(\nu_{\mathrm{obs}}) = \int_{\gamma} g^{3} j(\nu_{\mathrm{em}}) \, dl_{p}.
\end{equation}

The redshift factor in this case is represented by the equation $g = \nu_{\mathrm{obs}} / \nu_{\mathrm{em}}$, which establishes the relationship between the frequency of the observed photon $ \nu_{\mathrm{obs}}$ and the frequency of the emitted photon $\nu_{\mathrm{em}}$. In addition, $ j(\nu_{\mathrm{em}})$ is the emissivity per unit volume from the standpoint of a stationary emitter.   Here, the term $dl_{p}$ signifies an infinitesimal proper length, while $\gamma$ illustrates the path of the light ray. The redshift factor in the quintessence-like Fan-Wang BH is $g=f(r)^{1/2}$.  When the emission radial profile is $1/r^{2}$ \cite{2013PhRvD..87j7501B} and the emission is monochromatic with rest frame frequency $\nu_{t}$, the specific emissivity is expressed as follows
\begin{equation}
j(\nu_{\text{em}}) \propto \frac{\delta (\nu_{\text{em}} - \nu_t)}{r^2}.
\label{eq:emissivity}
\end{equation}

The proper length can be calculated through Eq. (\ref{metr_generic}), and it is expressed as follows
\begin{equation}
dl_p = \sqrt{\frac{dr^2}{f(r)} + r^2 d\varphi^2} = \sqrt{\frac{1}{f(r)} + r^2 \left( \frac{d\varphi}{dr} \right)^2} dr,
\label{eq:proper_length}
\end{equation}

Consequently, substituting Eq. (\ref{eq:photon_trajectory}) with Eq. (\ref{eq:proper_length}) allows us to deduce the specific intensity as perceived by a remote observer, which is expressed as
\begin{equation}
I(\nu_{\text{obs}}) = \int_{\gamma} \frac{f(r)}{r^2} \sqrt{1 + \frac{b^2 f(r)}{r^2 - b^2 f(r)}} dr.
\label{eq:observed_intensity}
\end{equation}

Eq. (\ref{eq:photon_trajectory}) illustrates that the trajectory of light rays is dictated by the impact parameter, denoted as $b$. In instances where $b=b_{ph}$, which corresponds to an energy $E = E_{ph}$, photons are induced to orbit the BH. For cases with $b>b_{ph}$, corresponding to an energy $E<E_{ph}$, the trajectory of light rays remains under the influence of the BH's gravitational field; however, the trajectories will not result in stable orbits. In such cases, they will be subjected to deflection by the gravitational field of the BH. Conversely, for $b<b_{ph}$, corresponding to an energy $E>E_{ph}$, the photon is characterized by surplus energy. Under these conditions, the photon is unable to overcome the gravitational attraction of the BH and will ultimately be drawn into the singularity of the BH. Consequently, to investigate the photon intensity related to the light trajectories, we depicted the observed intensity as a function of the impact parameter, denoted by $b$. 

\begin{figure*}[ht]
\begin{minipage}{0.49\linewidth}
\centering
\includegraphics[width=\linewidth]{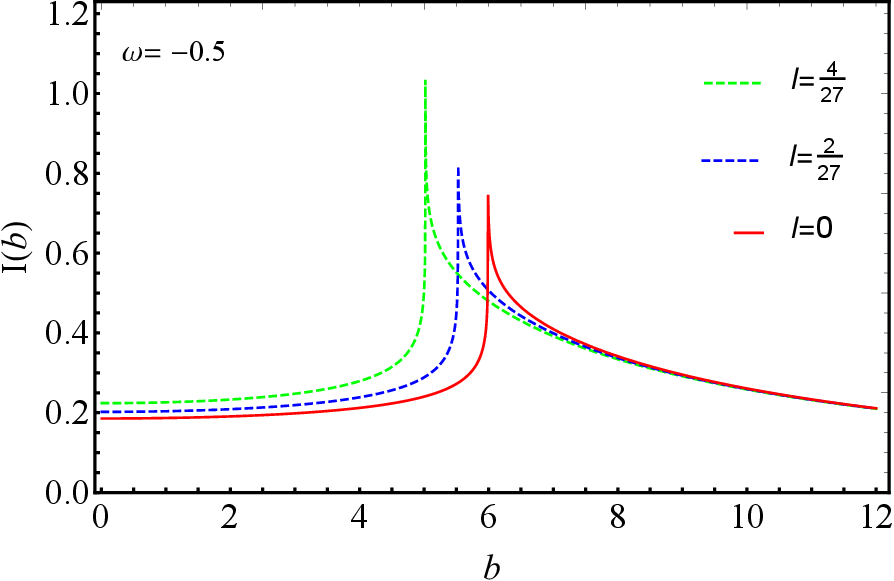}
\end{minipage}
\hfill 
\begin{minipage}{0.49\linewidth}
\centering
\includegraphics[width=\linewidth]{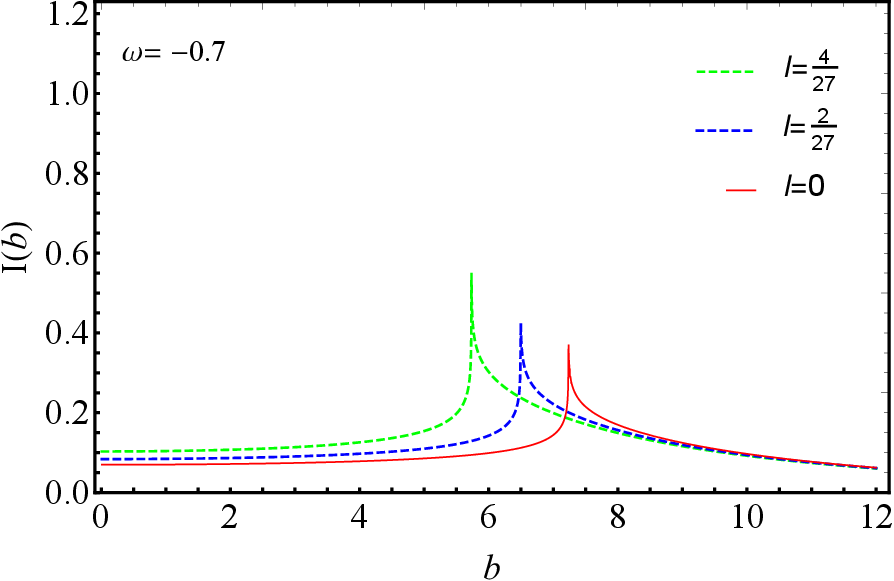}
\end{minipage}
\caption{The observed intensity $I(\nu_{\text{obs}})$ for static spherical accretion flow around a quintessence-like Fan-Wang and quintessence-like Schwarzschild BHs with $\omega=-0.5$ (left panel) and $\omega=-0.7$ (right panel) for $a = 0.05$ at $M = 1$.}
\label{fig:StaticIntens}
\end{figure*}

In Fig. \ref{fig:StaticIntens}, the graphical depictions illustrate the intensity as a function of the impact parameter $b$ for the quintessence-like Schwarzschild and quintessence-like Fan-Wang. The left panel corresponds to $\omega=-0.5$, while the right panel shows the case for $\omega=-0.7$. For a fixed value of $\omega=-0.5$, the peak intensity values for the quintessence-like Fan-Wang BH at $l=4/27$ and $l=2/27$  exceed those of the quintessence-like Schwarzschild BH.
In the case of $\omega=-0.7$, the peak intensity value for the quintessence-like Fan-Wang BH also exceeds that of the quintessence-like Schwarzschild BH for $l=0$. Furthermore, the peak intensity for quintessence-like Fan-Wang and quintessence-like Schwarzschild,  observed at $\omega=-0.7$ is lower than that at  $\omega=-0.5$.
As seen in both panels of Fig. \ref{fig:StaticIntens}, the intensity increases with the impact parameter $b$, reaching a peak at $b_{ph}$, and then rapidly decreases to lower values. The maximal intensity at $b=b_{ph}$ arises from the fact that photons executing multiple orbits around the BH as they approach the photon sphere. 
Additionally, at fixed $\omega$, the peak intensity increases with increasing $l$, indicating that the quintessence-like Schwarzschild BH with $l=0$ exhibits the lowest peak value of intensity. The appearance of cusp-like features in the intensity profiles is associated with the accumulation of light rays near the photon sphere. Photons with impact parameters close to the critical value spend more time in the strong gravitational field region, resulting in enhanced brightness and the formation of sharp peaks in the observed intensity. Furthermore, when comparing the intensity values for the same $l$, the peak value of intensity at $\omega=-0.7$ is lower than that at $\omega =-0.5$. This indicates that the observed intensity decreases in the presence of exotic anisotropic matter.

\begin{figure*}[ht]
\center{\includegraphics[width=0.3\linewidth]{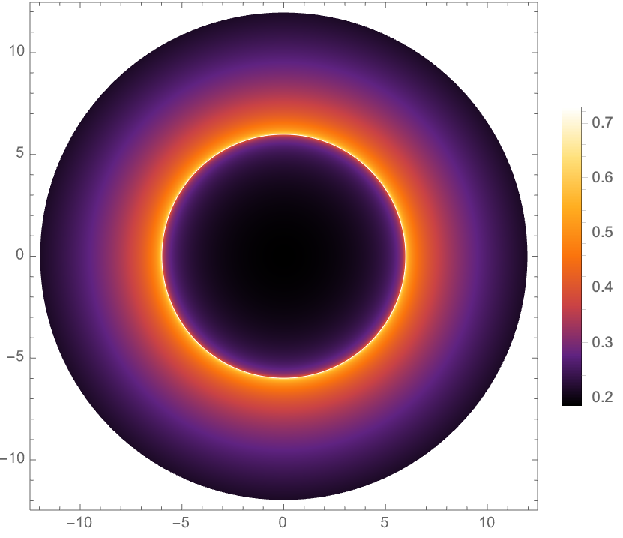}\quad \includegraphics[width=0.3\linewidth]{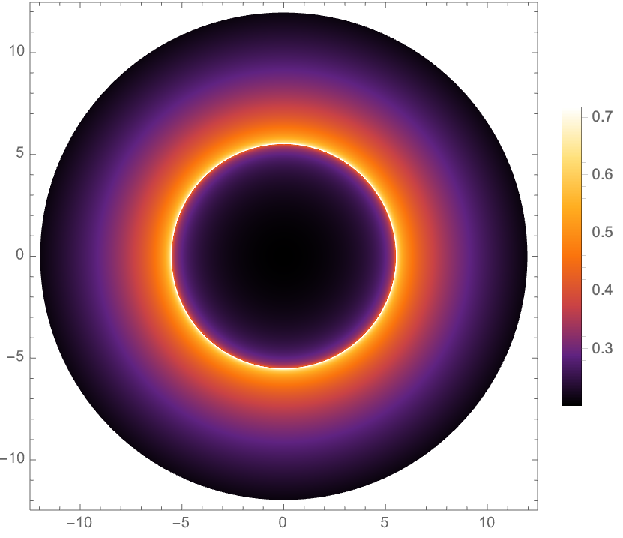}\quad\includegraphics[width=0.3\linewidth]{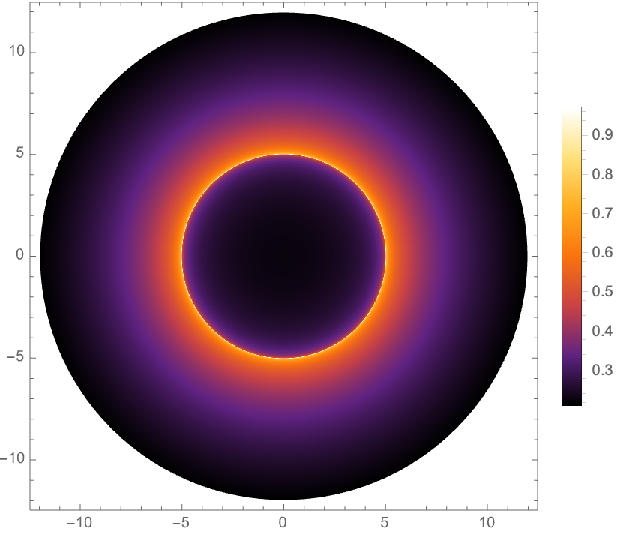} }\\
\center{\includegraphics[width=0.3\linewidth]{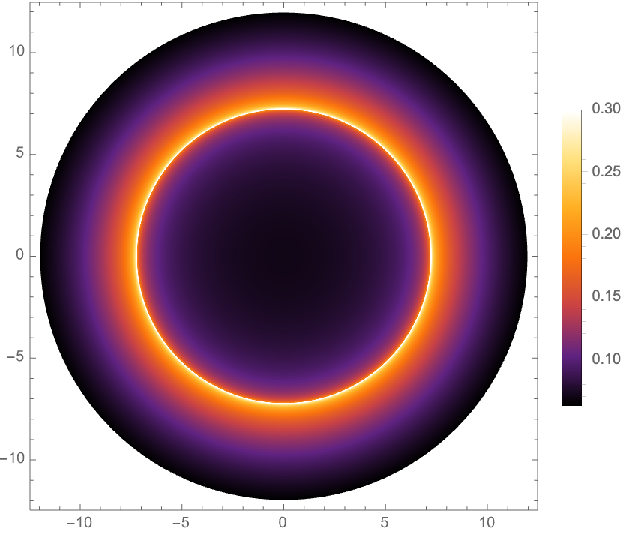}\quad \includegraphics[width=0.3\linewidth]{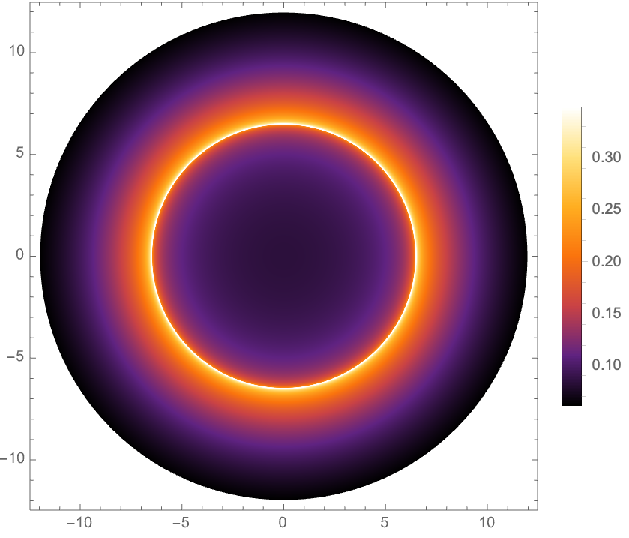}\quad \includegraphics[width=0.3\linewidth]{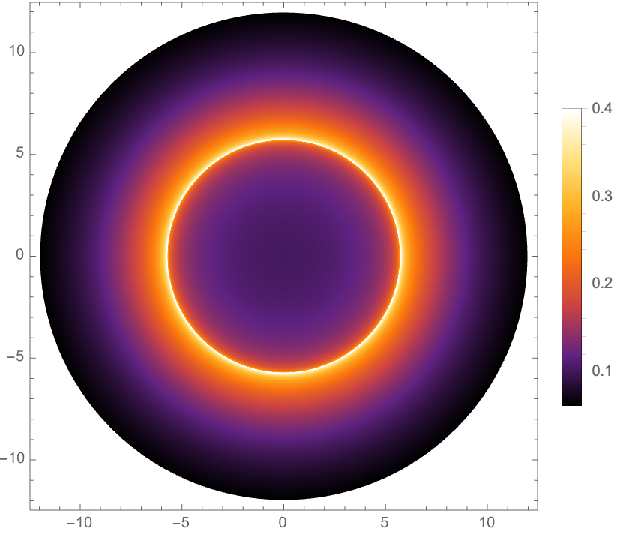}}
\caption{Shadows and photon rings for static spherical accretion flow with $\omega = - 0.5$ (top row) and $\omega = - 0.7$ (bottom row) for various values of $l$, $a = 0.05$ and $M = 1$. The parameter $l$ from left to right taking on values of $0$, $2/27$, and $4/27$, respectively.}
\label{fig:StatAcc0507}
\end{figure*}

Fig. \ref{fig:StatAcc0507} illustrates the shadows and photon rings of quintessence-like Fan-Wang BHs for various values of $l$ for cases $\omega=-0.5$ and $\omega=-0.7$. When $\omega$ is fixed, increasing the parameter $l$ results in a decrease in the radius of the shadow and photon rings of the quintessence-like Fan-Wang BH, while causing an increase in luminosity. It is notable that the quintessence-like Schwarzschild BHs, characterized by $l=0$ exhibits the lowest luminosity compared to the quintessence-like Fan-Wang BHs. This is attributed to the stronger gravitational field of the quintessence-like Schwarzschild BH with $l=0$, which results in an increased deflection of light and consequently more photons being captured by the BH. This phenomenon leads to the observation of lower luminosity of the shadow and photon ring compared to that of quintessence-like Fan-Wang BHs. Furthermore, a comparison between the plots in the top and bottom rows of the figure reveals the influence of the state parameter of the quintessence-like matter on the BH shadow and photon ring. For a fixed value of $l$, it is observed that the luminosity of the photon ring increases with the increasing absolute value of $\omega$. Furthermore, by analyzing the effective potential of test particles (photon), we have investigated the null geodesics and the kinds of orbits of the Hayward BH surrounded by quintessence-like corresponding to different energy levels.

\subsection{Infalling spherical accretion}

At present, our investigation is concentrated on optically thin accretion, which is hypothesized to result from the accretion of infalling matter. Since most accreting materials in the universe behave dynamically, this infalling model is thought to be more realistic than the static accretion model. Eq. (\ref{eq:observed_intensity}) remains employed in our examination of the shadow produced by the infalling accretion. Contrary to the static accretion model, the redshift factor in the context of infalling accretion is intricately associated with the velocity of the accreting material, and it is expressed by
\begin{equation}\label{g}
g = \frac{ k_{\alpha} u_{obs}^{\alpha} }{ k_{\beta} u_{em}^{\beta} },
\end{equation}
here $k^\mu \equiv \dot{x}^\mu$, $u_{\text{obs}}^\mu = (1, 0, 0, 0)$, and $u_{\text{em}}^\mu$ denote the photon four-velocity, the four-velocity of a distant observer, and the four-velocity associated with accretion, respectively. Eq. (\ref{eq:time_derivative}) indicates that $ k_{t} = 1/b $ is a constant. Furthermore, the value of $k_{r} $ can be derived from the condition that  $k_{\mu}k^{\mu} = 0$. Consequently, we derive
\begin{equation}
\frac{k_r}{k_t} = \pm\sqrt{\frac{1}{f(r)}\left(\frac{1}{f(r)}-\frac{b^2}{r^2}\right)},
\end{equation}
here the upper and lower signs correspond to the scenarios in which the photons are moving towards or away from the BH, respectively. The four-velocity of the infalling accretion is $\left(u_{\text{em}}^{t},\, u_{\text{em}}^{r},\, u_{\text{em}}^{\theta},\, u_{\text{em}}^{\varphi}\right) = \left(1/f(r),\, -\sqrt{1 - f(r)},\, 0,\, 0\right)$. By employing these equations, the redshift factor in Eq. (\ref{g}) can be determined as follows
\begin{equation}
g = \frac{1}{u_{\text{em}}^t + \left( \frac{k_r}{k_t} \right) u_{\text{em}}^r}.
\end{equation}

Furthermore, the proper distance is defined as
\begin{equation}
dl_p = k_\alpha u_{\text{em}}^\alpha ds = \frac{k_t}{g^3 \lvert k_r \rvert} dr,
\end{equation}
where $s$ denotes the affine parameter along the trajectory of the photon. By positing that the specific emissivity is monochromatic, the specific intensity $I(\nu_{obs}) $ for the scenario of infalling spherical accretion can therefore be expressed as
\begin{equation}
I = \int \frac{g^3}{r^2 \sqrt{\frac{1}{f(r)} \left( \frac{1}{f(r)} - \frac{b^2}{r^2} \right)}}dr.
\end{equation}

\begin{figure*}[ht]
\begin{minipage}{0.48\linewidth}
\centering
\includegraphics[width=\linewidth]{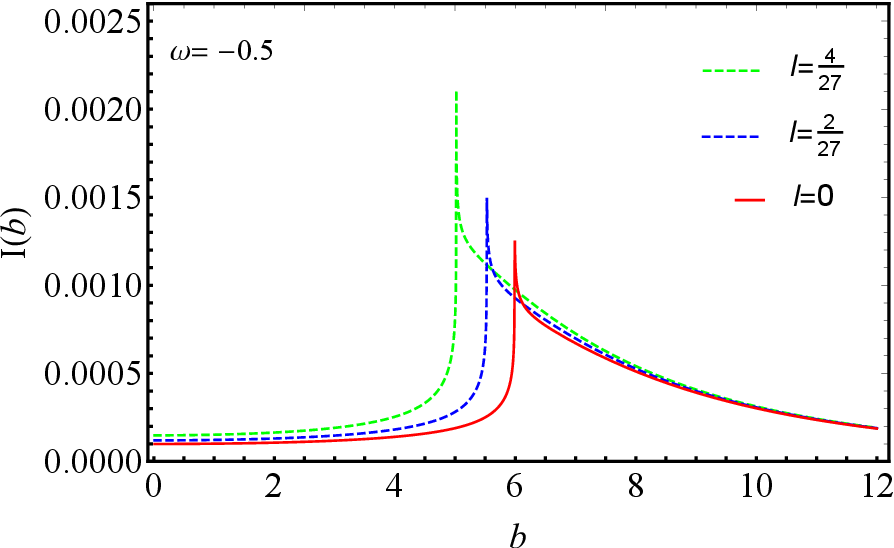}
\end{minipage}
\hfill 
\begin{minipage}{0.48\linewidth}
\centering
\includegraphics[width=\linewidth]{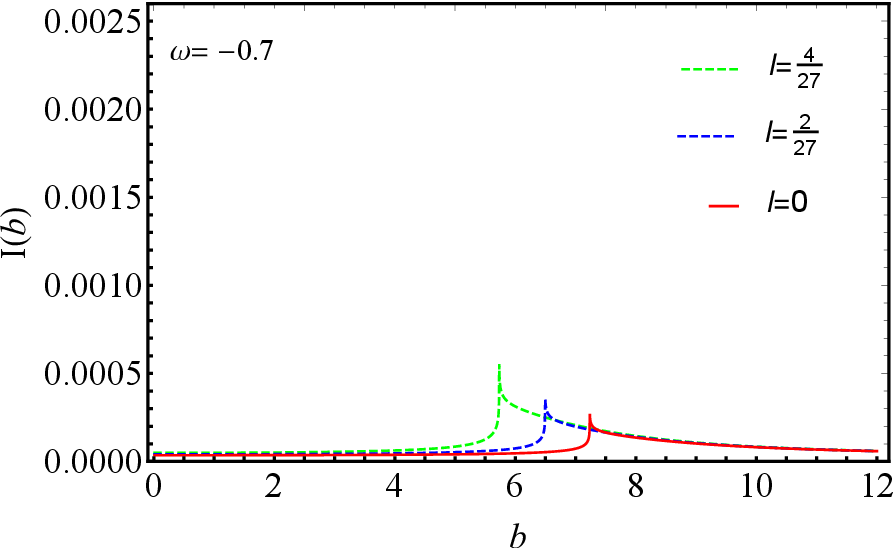}
\end{minipage}
\caption{The observed intensity $I(\nu_{\mathrm{obs}})$ for infalling spherical accretion flow around a  quintessence-like Fan-Wang and quintessence-like Schwarzschild BHs with $\omega=-0.5$ (left panel) and $\omega=-0.7$ (right panel) for $a = 0.05$ and $M = 1$.}
\label{fig:InfInten}
\end{figure*}

Fig. \ref{fig:InfInten} presents the observed intensity as a function of the parameter $b$, with $\omega=-0.5$ in the left panel and $\omega=-0.7$ in the right panel. As seen in the figure, the intensity is analogous to that of a static spherical accretion flow, with the peak occurring at $b_{ph}$, however, the maximum intensity in this case is lower than that observed in the static scenario.

\begin{figure*}[ht]
\center{\includegraphics[width=0.3\linewidth]{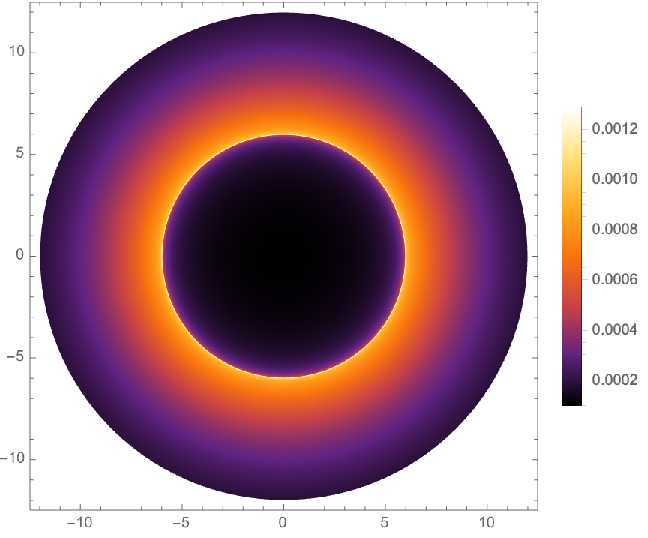}\quad \includegraphics[width=0.3\linewidth]{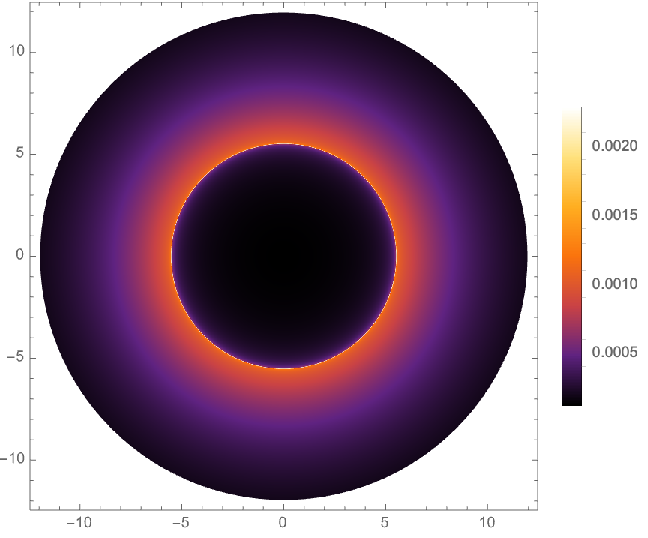}\quad\includegraphics[width=0.3\linewidth]{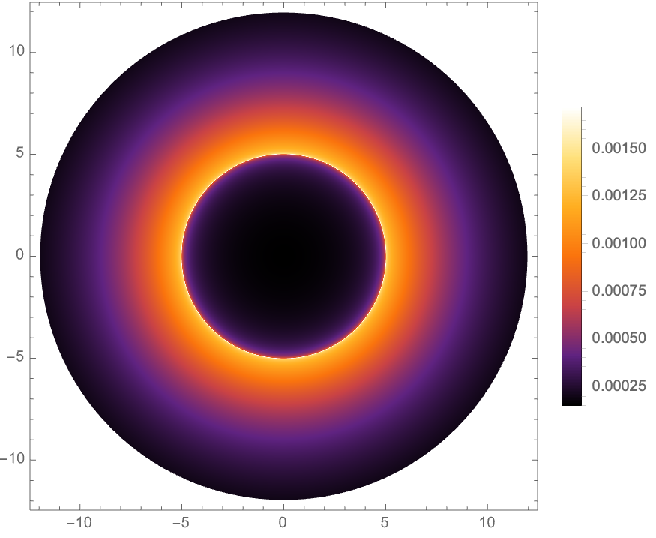} }\\
\center{\includegraphics[width=0.3\linewidth]{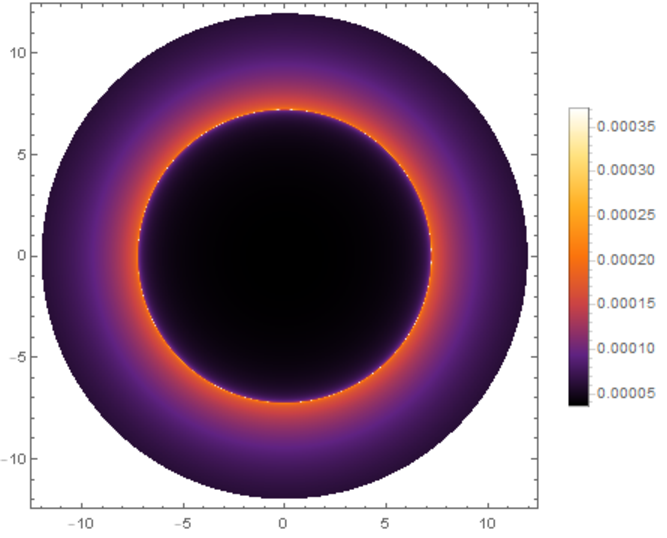}\quad \includegraphics[width=0.3\linewidth]{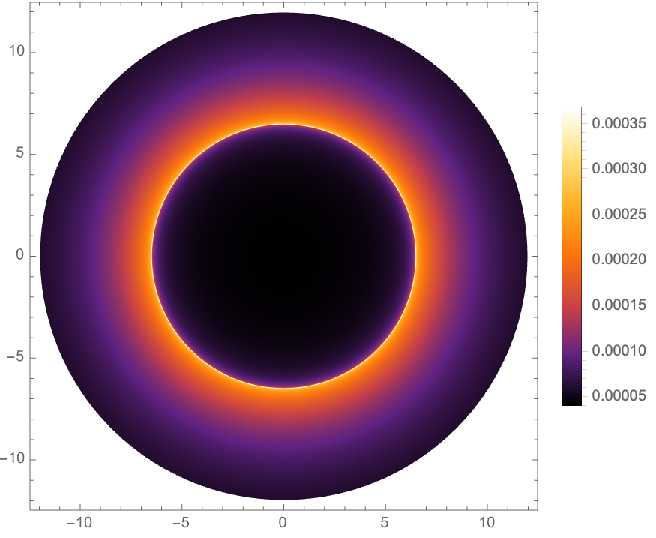}\quad \includegraphics[width=0.3\linewidth]{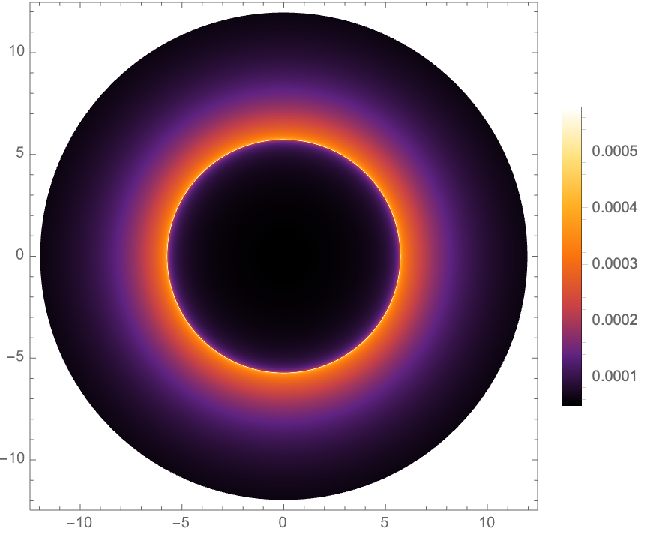}}
\caption{Shadows and photon rings for infalling spherical accretion flow with $\omega=-0.5$ (top row) and $\omega=-0.7$ (bottom row) for various values of $l$ (from left to right, $l=0$, $2/27$, and $4/27$), $a = 0.05$ and $M = 1$.}
\label{fig:InfAcc0507}
\end{figure*}

Fig. \ref{fig:InfAcc0507} illustrates the representation of the shadow’s image corresponding to the two values $\omega=-0.5$ and $\omega=-0.7$.The graphical depiction shows that, for a fixed value of $\omega$, both the photon rings and the luminosity of the shadow increase with the parameter $l$.

\section{Final outlooks and perspectives}\label{Conclusion}

Within the context of an external environment, we characterized the main properties of a class of regular solutions inspired by the Fan-Wang BH embedded in a matter field with anisotropic pressures, emulating quintessence as it exhibits constant and negative pressures, albeit the fluid \emph{is effectively not} quintessence, as it does not show pressure isotropy.

In this respect, we extended previous results, obtained by considering the Kiselev solution in a anisotropic (in pressure) field of matter, to RBHs belonging to the class of Fan-Wang BHs. 
More precisely, our  solution is described by three key parameters, namely the total mass of the source, $M$, the charge-related parameter, $l$, and the state parameter, $\omega$, indicating that the external environment mimics matter with pressure, violating the Zel'dovich limit as $\omega<0$. In addition to these terms, a further normalization constant, $a$, has been considered with the aim of examining the combined influence of these parameters on the BH structure and on  properties of null geodesics.

In addition, we conducted a detailed analysis of both static and infalling spherical accretion processes, exploring how the accretion flow affects the optical appearance of the BH. Numerical evaluations were performed to determine the radii of the inner horizon $r_{-}$, event horizon $r_{h}$, and cosmological horizon $r_{c}$, as well as the radius of the photon sphere $r_{ph}$ and its corresponding impact parameter $b_{ph}$, for various values of $l$ and $\omega$.

We further analyzed the effective potential experienced by test particles to study the behavior of null geodesics and the associated orbital trajectories around the Fan-Wang BH in the presence of quintessence-like matter, considering different energy levels.

Our findings indicated that the inclusion of our exotic fluid significantly modifies the observable features of the BH, such as the shadow size, the structure of the photon ring, and their respective luminosities, compared to the standard Fan-Wang BH. However, a comparison between Schwarzschild and Fan-Wang spacetimes, surrounded by the underlying external field, revealed only marginal differences in these optical characteristics.

Overall, this study may be generalized since the influence of cosmic fluids, with negative equations of state, can lead to evident experimental signatures to detect at  astrophysical level. 

Accordingly, we intend to study additional spacetimes, extending previous literature based on configurations in which regular solutions are embedded in exotic external fields, see e.g. \cite{2025arXiv250400332A,2025EPJP..140..476D,2025arXiv250703701M,2025arXiv250707372A,2025arXiv250318027A,2024EPJC...84..860L,2023PDU....4201311L,2024ResPh..5807499W,2023PDU....4201293H}. In this respect, we will investigate situations where the effects of negative pressure affects thermodynamic and optical properties around a given solution, modifying the equation of state evolution. Last but not least, we will explore how scalar, vector, or more complicated fields,  will determine the evolution of the underlying exotic fluid around a given regular solution, since, in this work, we  limited our attention to the purely barotropic case.

\begin{acknowledgments}
    YeK acknowledges Grant No. AP23488743, TK acknowledges Grant No. AP19174979, KB acknowledges Grant No. AP19680128 from the Science Committee of the Ministry of Science and Higher Education of the Republic of Kazakhstan. OL acknowledges the support by the Fondazione ICSC, Spoke 3 Astrophysics and Cosmos Observations. National Recovery and Resilience Plan (Piano Nazionale di Ripresa e Resilienza, PNRR) Project ID $CN00000013$ ``Italian Research Center on High-Performance Computing, Big Data and Quantum Computing" funded by MUR Missione 4 Componente 2 Investimento 1.4: Potenziamento strutture di ricerca e creazione di ``campioni nazionali di R\&S (M4C2-19 )" - Next Generation EU (NGEU).
    MM acknowledges the support of the European Union - NextGenerationEU, Mission 4, Component 2, under the Italian Ministry of University and Research (MUR) - Strengthening research structures and creation of "national R\&D champions" on some Key Enabling Technologies - grant CN00000033 - NBFC - CUPJ13C23000490006.
\end{acknowledgments}

\bibliography{0refs}
\end{document}